\documentclass[preprint,showpacs,preprintnumbers,amsmath,amssymb]{revtex4}

\usepackage{graphicx}
\usepackage{dcolumn}
\usepackage{bm}
\usepackage{latexsym}
\usepackage{amsfonts}
\usepackage{amssymb}
\usepackage{amsmath}
\usepackage{color}

\begin{document}

\preprint{DCPT-09/61, KUNS-2229}

\title{Cosmological Magnetic Fields from Inflation and Backreaction}

\author{Sugumi Kanno$^{1)}$}
\author{Jiro Soda$^{2)}$}
\author{Masa-aki Watanabe$^{2)}$}
\affiliation{1) Centre for Particle Theory, Department of Mathematical
Sciences, Durham University,
South Road, Durham, DH1 3LE, UK}
\affiliation{2) Department of Physics,  Kyoto University, Kyoto, 606-8501, 
Japan}

\date{\today}

\begin{abstract}
We study the backreaction problem in a mechanism of
magnetogenesis from inflation.
In  usual analysis, it has been assumed that the backreaction due
 to electromagnetic fields spoils inflation once it becomes important. 
 However, there exists no justification for this assumption. 
Hence, we analyze  magnetogenesis from inflation
by taking into account the backreaction. 
On the contrary to the naive expectation, we show that
  inflation still continues even after the backreaction 
begins to work.  Nevertheless, it turns out that  
creation of primordial magnetic fields is significantly suppressed 
due to the backreaction.
\end{abstract}

\maketitle

\section{Introduction}

The origin of cosmological magnetic fields still remains an enigma 
in modern precision cosmology~\cite{Grasso:2000wj,
Kronberg:2007dy}. In particular, how to make
magnetic fields on Mpc scales would be a big challenge. As an attractive possibility,
 creation of primordial magnetic fields from inflation 
 has been discussed~\cite{Turner:1987bw,Ratra:1991bn}. 
 In particular, 
 a nonminimal kinetic term of vector fields in supergravity 
can be used to generate primordial cosmological magnetic 
fields~\cite{Martin:2007ue}.

Recently, the backreaction issue in this mechanism is 
raised~\cite{Demozzi:2009fu}. There,
it is pointed out that magnetic fields from inflation may not be a viable
mechanism when the backreaction is taken into account. 
The difficulty to create magnetic fields from inflation originates in
the existence of electric fields whose energy density rapidly grows
and soon catches up with the energy density of  inflaton. 
Thus, the backreaction begins to work. 
The crucial assumption here is that
the backreaction destroys inflation. Hence, creation of magnetic fields
cannot be compatible with a sufficiently long inflation.
However, no one has proved the above assumption. 
Rather, there is a counter example that the backreaction of vector fields
does not spoil inflation~\cite{Watanabe:2009ct}.
Hence, it is worth studying magnetogenesis from inflation
 by taking into account the backreaction.

In this paper, we reconsider  generation of magnetic fields during inflation.
We take into account the backreaction of electromagnetic fields
in an inflationary scenario. It turns out that
 the resultant cosmological evolution with backreaction is drastically different
from the one without backreaction. Indeed, the dynamics of 
inflaton is quite different and the universe
is anisotropically inflating although anisotropy is very small~\cite{Watanabe:2009ct}. 
In this new background, we consider  generation of magnetic fields
 and conclude that cosmological magnetic fields could be hardly produced
during inflation due to the backreaction although inflation still continues
in spite of backreaction.

The organization of the paper is as follows.
In section II, we present our model and clarify the condition when
the backreaction becomes important. 
In section III, we review a naive perturbative analysis and
reveal the necessity of the backreaction.
In section IV, we take into account the backreaction and
show that the backreaction changes the dynamics of inflaton significantly.
In this new inflationary background, we study generation
of magnetic fields and calculate the power spectrum of magnetic fields.
We find that it is difficult to produce
magnetic fields from inflation due to  backreaction
although inflation still continues.
The final section is devoted to conclusion.

\section{Models}
\label{sc:basic}

In this section, we present our models and basic equations
for discussing the backreaction~\cite{Watanabe:2009ct}. 
Here, we envisage the chaotic inflation although it is
easy to extend analysis to other scenarios.
We consider  electromagnetic fields in a homogeneous anisotropic universe
and explain why it is believed that inflation
is spoiled when the backreaction of electromagnetic fields becomes important.
Then, treating electromagnetic fields perturbatively, we clarify
in which case the backreaction becomes important. 

Let us consider the following action for the gravitational field, the inflaton
 field $\phi$ and the electro-magnetic
vector field $A_\mu$ coupled with $\phi$:
\begin{eqnarray}
S=\int d^4x\sqrt{-g}\left[~\frac{M_p^2}{2}R
-\frac{1}{2}\left(\partial_\mu\phi\right)\left(\partial^{\mu}\phi\right)
-V(\phi)-\frac{1}{4} f^2 (\phi) F_{\mu\nu}F^{\mu\nu}  
~\right] \ ,
\label{action1}
\end{eqnarray}
where $M_p$ is the reduced Plack mass, $g$ is the determinant of the metric, $R$ is the
Ricci scalar, $V(\phi)$ is the inflaton potential, $f(\phi)$ is the coupling function of the inflaton field to the vector one, respectively.
 The field strength of the vector field is defined by 
$F_{\mu\nu}=\partial_\mu A_\nu -\partial_\nu A_\mu$. 

Thanks to the gauge invariance, we can choose the gauge $A_0 =0$.
 Without loss of generality,
we can take $x$-axis in the direction of the vector.
Hence, we take the homogeneous fields of the form
$
A_\mu=(~0,~A_x(t),~0,~0~)
$
and 
$ 
\phi=\phi(t) \ .
$
Note that we have assumed the direction of the vector field does
not change in time, for simplicity. 
This field configuration holds  plane symmetry in the plane 
perpendicular to the vector.
Then, we take the metric to be 
\begin{eqnarray}
ds^2=- dt^2+e^{2\alpha(t)}\left[~ 
e^{-4\sigma(t)}dx^2    
+e^{2\sigma(t)}\left( dy^2 + dz^2\right)~\right] \ ,
\label{metric}
\end{eqnarray}
where the cosmic time $t$ is used.
Here, $a \equiv e^\alpha$ is an isotropic scale factor and $\sigma$ represents
a deviation from the isotropy.  It should be noted that we need to consider
an anisotropic spacetime from consistency when we treat a background vector field.
 With above ansatz, 
one obtains the equation of motion for the vector field which is
easily solved as
\begin{eqnarray}
\dot{A_x} = f^{-2}(\phi ) e^{-\alpha -4\sigma}p_{A}, 
\label{eq:Ax}
\end{eqnarray}
where an overdot denotes a derivative with respect to the cosmic time $t$
and $p_A$ denotes a constant of integration. 
Substituting (\ref{eq:Ax}) into other equations, we obtain basic equations
\begin{eqnarray}
\dot{\alpha}^2  &=& \dot{\sigma}^2
+\frac{1}{3 M_p^2}\left[ \frac{1}{2} \dot{\phi}^2+V(\phi)
+\frac{p_{A}^2}{2}f^{-2} (\phi) e^{-4\alpha-4\sigma }  \right] \ , 
\label{hamiltonian}\\
\ddot{\alpha} &=& -3\dot{\alpha}^2 + \frac{1}{M_p^2} V(\phi )
 +\frac{ p_{A}^2}{6 M_p^2}f^{-2}(\phi )e^{-4\alpha -4\sigma}, 
\label{evolution:alpha}\\
\ddot{\sigma} &=& -3\dot{\alpha}\dot{\sigma} 
+ \frac{ p_{A}^2}{3 M_p^2}f^{-2}(\phi )e^{-4\alpha -4\sigma} 
\label{eq:sigma}, \\
\ddot{\phi} &=& -3\dot{\alpha}\dot{\phi} -V_\phi(\phi ) 
+ p_{A}^2 f^{-3}(\phi )f_\phi (\phi ) e^{-4\alpha -4\sigma } 
\label{eq:phi} \ ,
\end{eqnarray}
where $V_\phi$ denotes the derivative of $V$ with respect to $\phi$.

From Eq.(\ref{hamiltonian}), we see the effective potential
$
V_{\rm eff} = V + p_A^2 f^{-2} e^{-4\alpha -4\sigma}/2
$
determines the inflaton dynamics. As the second term is
coming from the vector contribution, we refer it to
the energy density of the vector. Let's check if inflation
occurs in this model. Using Eqs.(\ref{hamiltonian}) and 
(\ref{evolution:alpha}), equation for acceleration of 
the universe is given by
\begin{eqnarray}
 \frac{\ddot{a}}{a} = \ddot{\alpha} + \dot{\alpha}^2 
 = - 2\dot{\sigma}^2 -\frac{1}{3 M_p^2} \dot{\phi}^2 
   + \frac{1}{3 M_p^2} \left[ V - \frac{p_A^2}{2} f^{-2}
    e^{-4\alpha -4\sigma } \right]  \ .
\end{eqnarray}
It is easy to see that if the energy density of the vector is dominant
 inflation soon ends. 
 This is the reason why it is usually supposed that the backreaction spoils
 inflation. However, as we will see in section IV, the relevance of
 backreaction does not necessarily imply the energetic dominance of the 
 electromagnetic fields.

To see when the backreaction of electromagnetic fields becomes relevant,
we consider electromagnetic fields in the isotropic universe. 
From Eq.(\ref{hamiltonian}), it is apparent that the fate of 
the electric field depends on the behavior of 
coupling function $f(\phi)$. By considering the critical 
case $f(\phi ) \propto e^{-2\alpha}$ for which the energy
density of the vector field remains almost constant during 
inflation, we can determine the functional form of $f$ 
under the assumption of slow-roll inflation.
We can use conventional slow-roll equations
\begin{eqnarray}
    \dot{\alpha}^2  =  \frac{1}{3 M_p^2}V(\phi), \quad
    3\dot{\alpha}\dot{\phi} = -V_\phi (\phi ) \ .
    \label{slow1}
\end{eqnarray}
Then, we have an equation
$
    d\alpha / d\phi = \dot{\alpha} /\dot{\phi} 
    = -  V(\phi) / M_p^2 V_\phi (\phi )  \ . 
$
This can be easily integrated as
$ 
     \alpha =  - \int V/ M_p^2 V_\phi d\phi \ .
$
Here, we have absorbed a constant of integration into the definition of 
$\alpha$. Thus, we obtain
\begin{equation}
f = e^{-2\alpha} =  e^{\frac{2}{M_p^2} \int \frac{V}{V_\phi} d\phi }  \ .
\label{critical}
\end{equation}
For the polynomial potential $V\propto \phi^2 $, we have
$
f =  e^{  \phi ^2/2M_p^2} \ .
$
Let us consider more general cases  
by introducing a parameter $c$ in the following 
form~\cite{Martin:2007ue}:
\begin{equation}
f = e^{ \frac{c}{2 M_p^2}  \phi^2 }  =e^{-2c\alpha}
                    \propto a^{-2c} \ ,  \label{key}
\end{equation}
where we have used the relation
\begin{eqnarray}
   e^{ \frac{1}{2M_p^2} \phi ^2} = e^{-2\alpha} \ .
\end{eqnarray}
We notice that the energy density of the vector field during inflation
 would be negligible when $c < 1$ and remains constant when $c=1$. 
 While it grows when $c > 1$ and
the growth rate of the energy density of vector fields can be
calculated as
\begin{eqnarray}
   \frac{p_{A}^2}{2}f^{-2} (\phi) e^{-4\alpha }
  \propto  \exp\left( 4(c-1)\alpha \right)  \ .
  \label{growth}
\end{eqnarray}
Apparently, we need to take into account the backreaction
for $c>1$.    
 In the next section, we consider generation of magnetic fields in the 
 conventional inflationary background driven by an inflaton.
 There, we will recognize  that $c>1$ is necessary for creating 
 cosmological magnetic fields and therefore the backreaction 
 turns out to be important. 
    
\section{Magnetic Fields from Inflation}
\label{sc:review}

In this section, we review the arguments made in the paper
 by Demozzi et al.~\cite{Demozzi:2009fu}.

Let us consider the Maxwell fields 
\begin{eqnarray}
S=\int d^4x\sqrt{-g}\left[-\frac{1}{4} f^2 (\phi) F_{\mu\nu}F^{\mu\nu}  
~\right] \ ,
\label{action1}
\end{eqnarray}
where we assumed a general coupling function (\ref{key}).
We will consider the isotropic inflationary background
\begin{eqnarray}
ds^2 = a^2 (\eta) \left( -d\eta^2 + dx^2 + dy^2 + dz^2 \right) \ ,
\label{FRW}
\end{eqnarray}
where, for convenience, we switched to the conformal time $\eta$.
It is well known that the physical degrees are described by
 the transverse vector which
 can be written in Fourier space as
\begin{eqnarray}
  A_i^T ({\bf x},\eta)= \sum_{\sigma =1,2} \int  \frac{d^3 k}{(2\pi)^{3/2}}
  A_{{\bf k}}^\sigma (\eta) \epsilon_i^\sigma ({\bf k})  e^{i{\bf k}\cdot {\bf x}} \ ,
\end{eqnarray}
where $\sigma$ denotes degrees of polarization and 
 polarization vectors $\epsilon^\sigma_i ({\bf k})$ satisfy
the relation  
$
k_i \epsilon_i^\sigma ({\bf k}) =0
$
and 
$  
\epsilon_i^\sigma (-{\bf k}) \epsilon_i^{\rho} ({\bf k}) =\delta_{\sigma \rho} \ .
$
Assuming the isotropic FRW universe (\ref{FRW}), we can deduce the action
for physical modes
\begin{eqnarray}
 S = \frac{1}{2} \sum_{\sigma =1,2} \int d\eta d^3 k f^2 (\phi) 
 \left[ A^{\sigma\prime}_{\bf k} A^{\sigma\prime}_{-{\bf k}} 
                   -k^2 A_{\bf k}^\sigma  A_{-{\bf k}}^\sigma \right] \ ,
\end{eqnarray}
where a prime represents a derivative with respect to the conformal time.
Let us quantize the system by promoting $A_{\bf k}^\sigma$
and the conjugate momentum
$
  \pi_{\bf k}^\sigma = f(\phi)^2 A_{-{\bf k}}^{\sigma\prime}
$
to operators satisfying
$
  \left[ A_{\bf k}^\sigma ,  \pi_{{\bf k}'}^{\rho} \right] 
  = i \delta^{\sigma \rho} \delta({\bf k}-{\bf k}')\ ,
$
and others are zero. 
Using creation and annihilation operators satisfying
$
  \left[ a_{\bf k}^\sigma , a_{{\bf k}'}^{\rho \dagger}  \right] 
  = \delta^{\sigma \rho} \delta({\bf k}-{\bf k}')  \ ,
$
we can expand $ A_{\bf k}^\sigma$ as
\begin{eqnarray}
    A_{\bf k}^\sigma 
    = u_{\bf k} a_{\bf k}^\sigma +u_{\bf k}^* a_{-{\bf k}}^{\sigma\dagger} \ ,
\end{eqnarray}
where mode functions have to satisfy equations of motion
\begin{eqnarray}
   u_{\bf k}'' + 2\frac{f'}{f} u_{\bf k}' + k^2 u_{\bf k} =0
\end{eqnarray}
and the normalization conditions
$
     u_{\bf k} u_{\bf k}^{*\prime} - u_{\bf k}^{*} u_{\bf k}^{\prime}
      = i / f^2   \ .
$
From the correlation function of the vector field
\begin{eqnarray}
   <0| A_i ({\bf x}) A^i (0)|0>
   = \int \frac{dk}{k} \frac{k^3}{\pi^2 a^2} |u_{\bf k}|^2 e^{i{\bf k}\cdot {\bf x}}
   = \int \frac{dk}{k}  \delta^2_A (k,\eta) e^{i{\bf k}\cdot {\bf x}} \ ,
\end{eqnarray}
we can read off the power spectrum
\begin{eqnarray}
  \delta_A^2 (k,\eta) = 
  \frac{|u_{\bf k}|^2 k^3 }{\pi^2 a^2 }   \ . 
  \label{field}
\end{eqnarray}
Similarly, we can deduce the power spectrum of the magnetic fields 
$B_i = \epsilon_{ijk} F^{jk}/2$
\begin{eqnarray}
  \delta_B^2 (k,\eta) = 
  \frac{|u_{\bf k}|^2 k^5 }{\pi^2 a^4 }  \ ,
  \label{magnetic}
\end{eqnarray}
where we used the formula
\begin{eqnarray}
  B^i B_i = \frac{1}{2a^4} F_{ik} F_{ik} 
  = \frac{1}{a^4} \left( \partial_i A_k \partial_i A_k
               - \partial_k A_i \partial_i A_k \right) \ .
\end{eqnarray}
The vacuum state $|0>$ can be specified by
the initial condition for the positive frequency mode at 
a sufficiently past time $\eta$
\begin{eqnarray}
  u_{\bf k} (\eta) 
  = \frac{1}{f \sqrt{2 k}} e^{-ik \eta } \ .
  \label{u:initial}
\end{eqnarray}
The above mode function on subhorizon scales
 connects to the superhorizon solutions
\begin{eqnarray}
  u_{\bf k} (\eta ) = C_1  + C_2  \int \frac{d\eta}{f^2} \ ,
  \label{u:solution}
\end{eqnarray}
where $C_1$ and $C_2$ are constants of integration. 
For the coupling function, we take
\begin{eqnarray}
  f = \left(\frac{a}{a_f} \right)^{-2c_{\rm eff}} \ ,
\end{eqnarray}
where $c_{\rm eff}$ is a parameter and
$a_f$ is the scale factor at the end of inflation $\eta_f$. 
In the case of the conventional slow roll inflation, following (\ref{key}),
we have $c_{\rm eff} = c$. 
Then, the solution (\ref{u:solution}) becomes 
\begin{eqnarray}
  u_{\bf k} = C_1  + C_2 a^{4c -1} \ ,
  \label{u:new}
\end{eqnarray}
where we used the relation $d\eta = da /(H_I a^2)$
and rescaled constants of integration.
Here, the Hubble parameter during inflation $H_I$ is assumed to be constant.
There are two branches where magnetic fields can be created. 
For a negative $c$, there exists no backreaction problem
because the electric fields do not exist in this case. 
 For this case, however, the analysis is not reliable due to
 the strong coupling problem~\cite{Demozzi:2009fu}.  
  Hence, we will not consider this case hereafter. 
For a positive $c$, the second term of (\ref{u:new}) which depends on time
is a relevant one. 
By matching (\ref{u:initial}) and (\ref{u:new})
 at the horizon crossing $a_k H_I =k$, the constant $C_2$ can be determined.
Thus, we obtain
\begin{eqnarray}
  u_{\bf k} 
  = \frac{1}{f_k \sqrt{2k}}  \left( \frac{a}{a_k} \right)^{4c-1} 
  = \frac{1}{f_k \sqrt{2k}} \left( \frac{H_I a}{k} \right)^{4c-1} \ ,
  \label{result}
\end{eqnarray}
where we defined $f_k = (a_k /a_f)^{-2c}$. 
Substituting (\ref{result}) into (\ref{magnetic}), 
we obtain magnetic fields at the end of inflation $\eta_f$
as
\begin{eqnarray}
  \delta_B (\lambda_p , \eta_f ) 
  = \frac{H_I^2}{\sqrt{2} \pi} \left( \frac{\lambda_p}{H_I^{-1}}\right)^{2c -3} \ ,
  \label{final}
\end{eqnarray}
where $\lambda_p = a_f /k$ is the physical wavelength at the end of 
inflation corresponding to the comoving wavenumber $k$. 
To get the flat spectrum, we need $c=3/2$. 
Assuming the GUT scale inflation
$H_I \sim 10^{-6} M_p$, we get $\delta_B \sim 10^{-12} M_p^2$
at the end of inflation.
Since $1G\sim 10^{-20} {\rm GeV}^2$, this implies $\delta_B \sim 10^{46} G$
at the end of inflation.
If we assume the instantaneous reheating and
 take $(M_p H_I )^{1/2}$ as a reheating temperature,
we have 
\begin{eqnarray}
    \frac{a_0}{a_f} \sim \frac{(M_p H_I )^{1/2}}{T_0} \sim 10^{29} \ .
\end{eqnarray}
where we used the temperature $T_0 \sim 10^{-13} $ GeV at present $a_0$.
Taking into account the relation $\delta_B \propto 1/a^2$,
 we obtain $\delta_B (\lambda_p ,\eta_0) 
 \sim (a_f /a_0 )^2 \delta_B (\lambda_p ,\eta_0) \sim 10^{-12} G$ at present.
This is close to the expected value $10^{-9} $G from observations.
 In that case, however, 
 the energy density of the electromagnetic fields
 $\rho_{em}$ exceeds the energy of the inflaton.
In fact, from the formula for the energy-momentum tensor
\begin{eqnarray}
  T_{\mu\nu} = f^2 (\phi ) \left[ 
                    F_{\mu\alpha} F_{\nu}{}^\alpha 
                    -\frac{1}{4} g_{\mu\nu} F_{\alpha \beta}F^{\alpha\beta}
                    \right] \ ,
\end{eqnarray}
we can calculate the energy density for the electro-magnetic fields
\begin{eqnarray}
  \rho_{em} =- <0|T^0{}_0|0> 
  = \frac{1}{4\pi^2 a^4} 
  \int \frac{dk}{k} k^3 f^2 \left[ 
  |u_{\bf k}^{\prime}|^2  +  k^2  |u_{\bf k}|^2\right] \ ,
\end{eqnarray}
which can be estimated as
\begin{eqnarray}
  \rho_{em}  \sim  H_I^4 \left( \frac{a}{a_i} \right)^{4c-4}  \ ,
  \label{electric}
\end{eqnarray}
where $a_i$ is the beginning of inflation. 
Once the energy density of the electro-magnetic fields $\rho_{em}$
becomes comparable to the energy density of the inflaton $\rho_\phi$,
namely, $\rho_{em} \sim \rho_\phi \sim  M_p^2 H_I^2 $,
we need to consider the backreaction.
More precisely, after the time $a /a_i \sim 10^{6}$
 we need to consider the backreaction. 
Usually, it is believed that once the electromagnetic fields becomes dominant,
the backreaction spoils inflation.
However, in the next section,
we will see inflation continues opposed to the naive expectation.
This is because the backreaction affects the dynamics of inflaton
 before the energy density of the
electromagnetic fields dominates that of the inflaton. 
 Thus, it is worth reexamining  generation of magnetic fields
 from inflation with the backreaction. 

\section{Magnetic Fields from Inflation with backreaction}
\label{sc:coe}

We consider the potential $V(\phi ) = m^2\phi^2 /2$ 
in this section. Hence, we adopt the coupling function 
$f(\phi)=e^{c \phi^2 /2M_p^2}$. 
Since we are going to look into the situation where the electromagnetic
field is not negligible, it is natural to consider an anisotropic spacetime
with a coherent vector field. However,
as the energy density of the vector 
field should be subdominant during inflation, the anisotropy is 
negligible to the lowest order. 
Hence we ignore $\sigma$ in the basic equations and regard 
it perturbative quantity.

\subsection{Inflation with backreaction}
The inflaton dynamics described by Eqs.(\ref{hamiltonian}) and (\ref{eq:phi}) 
can be written as
\begin{eqnarray}
\dot{\alpha}^2 &=& 
 \frac{1}{3 M_p^2}\left[ \frac{1}{2} \dot{\phi}^2
+\frac{1}{2}m^2\phi^2+\frac{1}{2}e^{-\frac{c}{M_p^2} \phi^2-4\alpha } p_{A}^2 
              \right]  \ , \label{h3} \\
\ddot{\phi} &=& -3\dot{\alpha}\dot{\phi} -m^2\phi 
+ \frac{c}{M_p^2} \phi e^{-\frac{c}{M_p^2}
                    \phi^2-4\alpha  }p_{A}^2\label{eq:phi3}      \ .        
\end{eqnarray}
When the effect of the vector field is comparable with that of the inflaton
field as source terms in (\ref{eq:phi3}), we get the relation 
$c p_A^2 e^{-c \phi^2/M_p^2 -4\alpha }/M_p^2 \sim m^2 $.
If we define the ratio of the energy density of the vector field 
$\rho_A\equiv p_A^2 e^{-c \phi^2/M_p^2 -4\alpha } /2$ to that of 
the inflaton $\rho_\phi\equiv m^2 \phi^2/2$ as
\begin{equation}
{\cal R} \equiv \frac{\rho_A}{\rho_\phi}
= \frac{p_{A}^2 e^{-\frac{c}{M_p^2} \phi^2-4\alpha}}{m^2\phi^2} \ ,
\label{R}
\end{equation}
we find the ratio becomes
${\cal R} \sim M_p^2 /c \phi^2$
when the above relation holds.
Since inflation takes place
typically at $ \phi \sim {\cal O} (10) M_p $, the ratio goes
${\cal R} \sim 10^{-2}$. Thus we find that the effect of the
vector filed in (\ref{h3}) is negligible even when it is comparable
with that of the scalar field in (\ref{eq:phi3}).

We notice that the above situation is not transient one but
an attractor.
Suppose that $\rho_A$ is initially negligible, 
${\cal R}_i \ll 10^{-2} $. In the conventional slow-roll 
inflationary phase (\ref{slow1}), 
the relation
$e^{-\phi^2/M_p^2} \propto e^{4\alpha} $ 
holds as was shown in (\ref{critical}).
Hence, the ratio ${\cal R}$ varies as
$
{\cal R} \propto e^{4(c-1)\alpha}.
$
As we now consider $c>1$, $\rho_A$ increases rapidly during inflation
and eventually reaches  ${\cal R} \sim 10^{-2}$.
Whereas, if the ratio is initially ${\cal R}_i \gg 10^{-2} $,
the inflaton climbs up the potential due to the effect
of the vector field in (\ref{eq:phi3}), hence $\rho_A$ will decrease rapidly
and again  ${\cal R} \sim 10^{-2}$ will be realized soon. 
Thus irrespective of initial conditions, $\rho_A$ will track $\rho_{\phi}$.

From these arguments, the inflaton dynamics after tracking is
governed by the modified slow-roll equations
\begin{eqnarray}
 \dot{\alpha}^2 &=& \frac{1}{6 M_p^2} m^2 \phi^2   \ , 
\label{h4}\\
3\dot{\alpha} \dot{\phi} &=& 
-m^2\phi+ \frac{c}{M_p^2} \phi p_{A}^2 e^{-\frac{c}{M_p^2} \phi^2-4\alpha  } \ . 
\label{eq:balance}
\end{eqnarray}
We refer to the phase governed by the above equations as the
second inflationary phase, compared to the first one
governed by the equations (\ref{slow1}). 
Using above equations, we can deduce
\begin{equation}
  \phi \frac{d\phi}{d\alpha}  = - 2 M_p^2 + \frac{2cp_A^2}{m^2}
                e^{-\frac{c}{M_p^2} \phi^2 -4 \alpha}  \ . \label{phi:alpha}
\end{equation}
This can be integrated as
$
  e^{-c \phi^2/M_p^2 -4\alpha} 
  = m^2 M_p^2 (c-1)/ c^2 p_A^2 \left[1+D e^{-4(c-1)\alpha} \right]^{-1}  ,
$
where $D$ is a constant of integration. This solution rapidly converges 
to 
\begin{eqnarray}
  e^{-\frac{c}{M_p^2} \phi^2 -4\alpha} 
                = \frac{m^2 M_p^2 (c-1)}{c^2  p_A^2} \ .
\label{attractor}
\end{eqnarray}
Thus, we found $\rho_A$ becomes constant during the second inflationary 
phase~\cite{Watanabe:2009ct}. In particular, from (\ref{attractor}),
 the relation
\begin{eqnarray}
   f =  e^{\frac{c}{2 M_p^2} \phi^2 } \propto a^{-2} 
\end{eqnarray}
holds in this phase. Thus, when the backreaction becomes relevant, 
 we effectively  have $c_{\rm eff}=1$. 

We note that the backreaction makes the expansion anisotropic.
In fact, as is done in a previous work~\cite{Watanabe:2009ct}, 
we can obtain a remarkable result from (\ref{eq:sigma}) as 
\begin{equation}
 \frac{\Sigma}{H_I} \equiv \frac{\dot{\sigma}}{\dot{\alpha}}
 = \frac{1}{3}\frac{c-1}{c} \epsilon \ ,
\end{equation}
where $\epsilon $ is a slow roll parameter and $\Sigma /H$
is the anisotropic expansion rate normalized by the Hubble parameter.
Typically, we have a small anisotropy $\Sigma/H \sim 10^{-2}$.
However, it is not impossible to detect this small number
by PLANCK~\cite{Pullen:2007tu}. 

\subsection{Generation of Magnetic fields}

In the situation where magnetic fields have a scale free spectrum $c=3/2$,
the formula (\ref{field}) tells us that 
the electric fields have a red spectrum. Hence, the largest scale
has a dominant contribution to the energy density.
We can assume the coherent electric fields with a definite direction
dominate the energy density of the electromagnetic fields.
Then, the result in the previous subsection is applicable. 

Now, we will take into account the backreaction.
The point is as follows.
Before the backreaction becomes important, 
we have the relation
\begin{eqnarray}
     f \propto \left(\frac{a}{a_f} \right)^{-2c}  \ .
     \label{before}
\end{eqnarray}
However, once the backreaction becomes important, we have
an attractor behavior 
\begin{eqnarray}
     f \propto \left(\frac{a}{a_f} \right)^{-2} \ .
     \label{after}
\end{eqnarray}
That means the effective $c_{\rm eff} $ changes from
$c_{\rm eff} =c$ to the critical value $c_{\rm eff}=1$
due to backreaction. Since we have the formula (see Eq.(\ref{electric}))
\begin{eqnarray}
  \rho_{em}  = 
   H_I^4 \left( \frac{a_b}{a_i} \right)^{4c-4} \ , 
\end{eqnarray}
the transition point $a_b$ occurs at $\rho_{em} \sim \rho_A \sim 10^{-2} \rho_\phi 
\sim 10^{-2} M_p^2 H_I^2 $, i.e.
\begin{eqnarray}
    \left( \frac{a_b}{a_i} \right)^{4c-4} 
    = 10^{-2}  H_I^{-2} M_p^2 \ .
\end{eqnarray}

We are now in a position to calculate the power spectrum of magnetic fields. 
First, we consider the modes which exit the horizon before $a_b$.
The superhorizon evolution of the mode function before $a_b$ is given by
\begin{eqnarray}
   u_{\bf k} (\eta) 
  = \frac{1}{f_k \sqrt{2k}}  \left( \frac{a}{a_k} \right)^{4c-1} \ ,
  \label{u:ev}
\end{eqnarray}
where we should note $f_k$ is defined by $f_k = (a_b /a_f)^{2c-2}(a_k /a_f)^{-2c}$
from the continuity at $a_b$. 
Since the evolution after $a_b$ becomes $u_{\bf k} \propto a^3$,
we obtain the mode function after $a_b$ as
\begin{eqnarray}
  u_{\bf k} 
  = \frac{1}{f_k \sqrt{2k}}  \left( \frac{a_b}{a_k} \right)^{4c-1} 
               \left( \frac{a}{a_b} \right)^3 \ .
\end{eqnarray}
From the formula (\ref{magnetic}), we obtain magnetic fields
\begin{eqnarray}
  \delta_B (\lambda_p , \eta_f ) 
  = \frac{H_I^2}{\sqrt{2} \pi} \left( \frac{\lambda_p}{H_I^{-1}}\right)^{2c -3}
            \left( \frac{a_b}{a_f} \right)^{2c-2}  \ .
            \label{main}
\end{eqnarray}
Compared to the cases with no backreaction (\ref{final}), the amplitude is
reduced by the factor
\begin{eqnarray}
   \left( \frac{a_b}{a_f} \right)^{2c-2}
   = 10^{-1}H_I^{-1} M_p \left( \frac{a_i}{a_f} \right)^{2c-2} \ .
\end{eqnarray}
For the flat spectrum $c=3/2$, we can deduce magnetic fields at the end of
inflation as
\begin{eqnarray}
   \delta_B (\lambda_p , \eta_f ) 
  =  10^{-1}H_I^{-1} M_p \left( \frac{a_i}{a_f} \right)
    \frac{H_I^2}{\sqrt{2} \pi}         \ .
\end{eqnarray}
Without backreaction, we anticipated $10^{-12}$G for the scale invariant case $c=3/2$.
However, by taking into account the backreaction, we 
have a suppression factor $a_b /a_f$ in (\ref{main}) which is  about $10^{-24}$.
Hence, we can expect at most $10^{-36}$G on Mpc scales at present.  
For modes which exit the horizon after the transition time $a_b$, 
by setting $c=1$ in (\ref{u:ev}), we obtain
\begin{eqnarray}
u_{\bf k} (\eta) 
  = \frac{1}{f_k \sqrt{2k}}  \left( \frac{a}{a_k} \right)^3 \ ,
\end{eqnarray}
where $f_k$ is now defined by $f_k = (a_k /a_f)^{-2}$.
Thus, we can calculate magnetic fields at the end of inflation as
\begin{eqnarray}
 \delta_B (\lambda_p , \eta_f ) 
  =   \frac{H_I^2}{ \sqrt{2}\pi} \left( \frac{\lambda_p}{H_I^{-1}}\right)^{-1}
             \ .
\end{eqnarray}
\begin{figure}[ht]
\includegraphics[height=7cm, width=9.5cm]{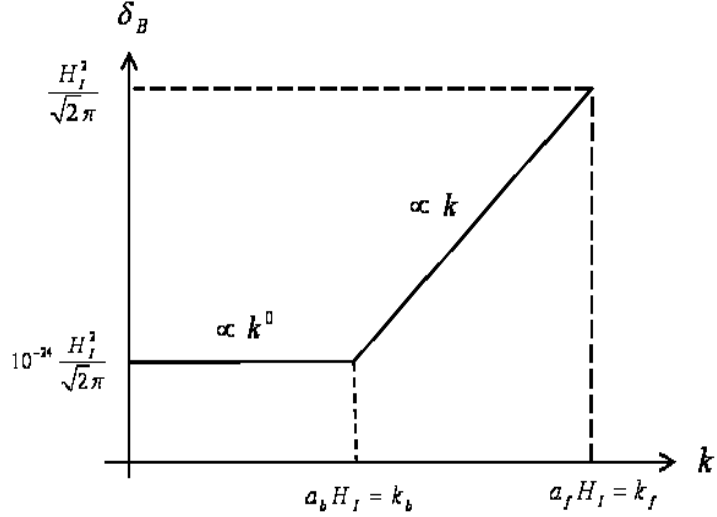}
\caption{The magnitude of magnetic fields is schematically 
depicted as a function of wavenumber $k$ for the case $c=3/2$.
There exists a break at $k_b = a_b H_I $ in the spectrum due to the backreaction.
As can be seen, the amplitude of magnetic fields on Mpc scales gets
 a suppression by $10^{-24}$ due to this break. }
\label{fig:cond}
\end{figure}
The resultant spectrum for $c=3/2$ is schematically depicted in Fig.1.
As is expected, we have a flat spectrum on large scales before the backreaction
becomes relevant. However, once the backreaction becomes important,
the spectrum becomes blue. 
Thus, after taking into account the backreaction, we realized that 
 primordial magnetic fields on large scales from inflation cannot be expected.

\section{Conclusion}

We studied cosmological magnetic fields in an inflationary scenario.
We have explained how the backreaction works in magnetogenesis
in the inflationary scenario. In usual analysis, it has been assumed
that the backreaction spoils inflation. However, there is no
justification for this assumption. Hence, we have incorporated the 
backreaction into the analysis and reanalyzed the production
process of magnetic fields from inflation. 
As a consequence, we have shown that
  inflation still continues after the backreaction 
becomes important. In spite of this fact, it turned out that creation of primordial 
magnetic fields is still significantly suppressed due to the backreaction.

The main point is that the energy density of
electric fields have to grow in order to have sufficient amplitude for
magnetic fields, which causes the backreaction. What we have found is
that the backreaction changes the dynamics of inflaton
and, as a consequence, makes the genesis of magnetic fields difficult. 
However,  inflation is not spoiled by the backreaction. 
Instead, an anisotropic inflationary
universe has been created due to backreaction. 
Although magnetic fields are negligible, since
 the expansion of the universe is anisotropic, other interesting phenomenology
such as the statistical anisotropy of primordial curvature perturbations
can be expected~\cite{Yokoyama:2008xw}. 
The possibility is now under investigation.

\begin{acknowledgements}
SK is supported by an STFC rolling grant.
JS is supported by the Japan-U.K. Research Cooperative Program, 
Grant-in-Aid for  Scientific Research Fund of the Ministry of 
Education, Science and Culture of Japan No.18540262,
Grant-in-Aid for  Scientific Research on Innovative Area
and the Grant-in-Aid for the Global COE Program 
``The Next Generation of Physics, Spun from Universality and Emergence".
\end{acknowledgements}

\end{document}